\pdfoutput=1
\documentclass[14pt]{article}
\usepackage{graphicx}
\usepackage[lmargin=1in, rmargin=0.8in,tmargin=2cm,bmargin=2.50cm]{geometry}
\usepackage[T1]{fontenc}
\usepackage[utf8]{inputenc}
\usepackage{authblk}
\usepackage{xcolor} 
\usepackage{booktabs}

\usepackage{pdfpages}

\begin{document}

\title{\bf{\vspace*{7\baselineskip}
A Computational Method for Modeling Arbitrary Junctions Employing Different Surface Integral Equation Formulations for Three-Dimensional Scattering and Radiation Problems}}
\author[1]{Hip\'olito G\'omez-Sousa}
\author[1]{\'Oscar Rubi\~nos-L\'opez}
\author[2]{Jos\'e \'Angel Mart\'inez-Lorenzo}
\author[1]{Marcos~Arias-Acu\~na}
\affil[1]{Department of Signal Theory and Communications, University of Vigo,\protect\\EI de Telecomunicaci\'{o}n, ES~36310 Vigo, Spain.\protect\\{\color{blue}{\{hgomez, oscar, marcos\}@com.uvigo.es}}}
\affil[2]{Northeastern University, 360 Huntington Ave.,\protect\\Suite 302 Stearns Center, Boston, MA~02115, USA.\protect\\{\color{blue}jmartinez@coe.neu.edu}}
\renewcommand\Authands{ and }
\date{}
\maketitle



\section*{Supplementary material}

# A computational method for modeling arbitrary junctions employing different surface integral equation formulations for three-dimensional scattering and radiation problems

H. Gómez-Sousa[a], Ó. Rubiños-López[a*], J. Á. Martínez-Lorenzo[b*] and M. Arias-Acuña[a]

[a]*Dept. of Signal Theory and Communications, University of Vigo, EI de Telecomunicación, ES 36310 Vigo, Spain*; [b]*Northeastern University, 360 Huntington Ave., Suite 302 Stearns Center, Boston, MA 02115, USA*

This paper presents a new method, based on the well-known method of moments (MoM), for the numerical electromagnetic analysis of scattering and radiation from metallic or dielectric structures, or both structure types in the same simulation, that are in contact with other metallic or dielectric structures. The proposed method for solving the MoM junction problem consists of two separate algorithms, one of which comprises a generalization for bodies in contact of the surface integral equation (SIE) formulations. Unlike some other published SIE generalizations in the field of computational electromagnetics, this generalization does not require duplicating unknowns on the dielectric separation surfaces. Additionally, this generalization is applicable to any ordinary single-scatterer SIE formulations employed as baseline. The other algorithm deals with enforcing boundary conditions and Kirchhoff's Law, relating the surface current flow across a junction edge. Two important features inherent to this latter algorithm consist of a mathematically compact description in matrix form, and, importantly from a software engineering point of view, an easy implementation in existing MoM codes which makes the debugging process more comprehensible. A practical example involving a real grounded monopole antenna for airplane-satellite communication is analyzed for validation purposes by comparing with precise measurements covering different electrical sizes.

## 1. Introduction

Electromagnetic scattering and radiation analysis from composite structures involving homogeneous dielectric and metallic materials has become an important problem in areas such as microwave systems engineering, antenna design and radar technology [1, 2]. Several approaches based on differential-equations formulations, which include FEM (finite element method) or FDTD (finite difference time domain) method, have been developed to computationally study the problem; however, as they are volumetric formulations strongly burdened with the discretization of the structure and the surrounding space, the applicability of these methods is very limited. Other alternative approaches have been successfully used for rigorously modeling this problem. In particular, when the

---

*Corresponding author. Email: oscar@com.uvigo.es

*Corresponding author. Email: jmartinez@coe.neu.edu

materials are piecewise homogeneous, surface integral equation (SIE) formulations, discretized by the well-known method of moments (MoM), are preferred. With the SIE-MoM formulation, the problem can be formulated in terms of surface integral equation over the conducting and dielectric surfaces and interfaces only, avoiding the discretization of volumes and thus reducing drastically the number of unknowns. However, the modeling of junctions formed by joining two or more distinct material surfaces each satisfying different boundary conditions remains a challenging problem.

The implementation difficulty of junction methods relies on that not only the typically-used boundary conditions for disconnected bodies, relating the inner and outer currents on a given surface, must be imposed. The implementation of this kind of methods also requires Kirchhoff's Law, relating the surface current flow in each region across a junction edge, as explained for instance in Section III of [3]. A junction edge is defined as the geometric place where two or more surface meshes connect at their common geometric edges, as indicated in Fig. 1.

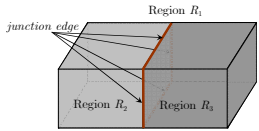

Figure 1. Junction edge for two cuboids in contact is indicated by arrows and highlighted in red. A junction edge is the geometric place where two or more surfaces connect at their common geometric edges. Note that a junction edge may also be composed of non-straight mesh edges.

Hereinafter, two different aspects of previously published junction methods are presented. Let us first explain how the continuity of surface currents, namely Kirchhoff's Law and boundary conditions at junction edges, is addressed in previous publications and how our algorithm for current continuity compares with these previous works. Later on in this introduction we discuss the particular features of previous implementations of generalized SIE formulations for multiple bodies. Each of these SIE generalizations must be employed together with a corresponding algorithm for current continuity at junction edges.

Some previous algorithms for current continuity described in [3–5] require the use of special half-basis functions at junction edges, instead of the ordinary full Rao-Wilton-Glisson (RWG) functions [6] which are preferable due to their simplicity. These special functions need different integration rules in different parts of each surface, which adds an additional burden in implementation.

A thorough explanation using full RWG functions for current continuity at junction edges is presented in [7]. Rigorous implementation rules are described in [7] in a general qualitative manner. This approach is very useful and informative from an instructive point of view; however, a detailed mathematical algorithmic description is missing, thus burdening the implementation of the described rules in real codes.

In paper [8], geometrical modeling and current continuity is performed by isoparametric surfaces, and surface currents at junctions are expressed as combinations, called multiplets, of original basis functions. In this algorithm in [8], there is no immediate way to perform the required MoM testing with the multiplets, instead of the ordinary RWG functions. This approach embraces many advantages –described in [8]– when developing new codes from the beginning, but hinders its implementation in existing codes which employ the simple commonly-used RWG bases.

Another approach, using ordinary full basis functions, is detailed in [9] in very methodical mathematical terms. However, the specific questions of modeling Kirchhoff's Law for enhanced field calculation, and the testing procedure at junction edges, are omitted in the explanations. Moreover, the relevant case involving a general dielectric-PEC (per-

fect electric conductor) junction is not taken into account as only penetrable junctions are addressed in [9]. Recently, some authors in [9] and a new author have completed in reference [10] the method in [9] by including in the mathematical description, amongst other important additions, Kirchhoff's Law. This completion is, according to a statement in [10], similar to the junction treatment in the reference [8] hereinabove commented. Even though a complete dielectric-PEC junction case was simulated for [10], the detailed mathematical description of this important special kind of problems was not yet included by the authors.

Finally, our algorithm for applying current continuity is similar to that presented in [12], Section 9.3.2. The main difference between both approaches is that the method in [12] operates on real RWG functions for surfaces in contact in a body mesh approach (using meshes of full bodies in contact), whereas the presented method only operates on fictitious RWG functions in a surface mesh approach (using independent meshes for each surface separating two adjacent scatterers, i.e., a different mesh for each separation surface instead of a mesh for each scatterer body). As a result, the approach in [12] requires that the meshing software, or the MoM code itself, marks out for every scatterer those triangles belonging to the surfaces in contact. A detailed description for general dielectric-PEC problems is also omitted in [12]; nevertheless, the topics covered in [12] regarding general dielectric junctions are described in both descriptive and mathematical manners with detailed and exhaustive information.

In this manuscript, a new simple algorithm for current continuity at junction edges, which leverages on traditional SIE formulations, is presented. The proposed continuity algorithm employs full RWG functions and, in contrast to some previous approaches, is applicable to penetrable junctions as well as to general PEC-dielectric junctions.

Just as in the previously commented references, the method proposed in this paper requires that the MoM matrix calculation part of the code executes a SIE formulation generalization for multibody problems. The SIE generalization to be used together with our current-continuity algorithm is developed in the Appendix of this manuscript. Even though composite scattering problems are largely discussed in the literature, our generalized SIE approach applicable to any ordinary SIE formulation is itself novel. Previous SIE generalization approaches for junctions, unlike ours, have the following issues: only generalize particular SIE formulations [3–5, 7, 8, 12], rely on duplication of unknowns [9] (unnecessarily raising the simulation time), and do not allow decoupling the enforcement of Kirchhoff's Law from the SIE generalization itself [3–5, 8, 10], thus hampering debugging and avoiding some appropriate software engineering practices. The need of such practices (writing reusable code –required due to limited development times and restricted number of software developers–, evolutionary software design, modular programming, etc.) arises from developing computational electromagnetic codes in micro teams [11], as can be usually the case in university research groups in Computational Electromagnetics.

The SIE generalizations and the current-continuity algorithm in this paper provide, together, a junction method which can be easily implemented in already-developed MoM codes capable of handling multiple disconnected scatterers. This easy implementation is achievable because *i)* the generalized SIE formulation part described in the Appendix requires only minor changes in codes for disconnected scatterers and *ii)* the computational algorithm in Section 2 for enforcing Kirchhoff's Law and boundary conditions at junction edges can be implemented independently from the rest of the MoM code. This easy implementation is due to the fact that the boundary conditions are enforced after filling the coefficients in the MoM matrix, and no special treatment is required for bodies in contact during the filling process. This implementation, independent from the rest of the MoM code, is a remarkable aspect from a software engineering point of view, as it

makes the debugging process and the addition of new features to the code much quicker and easier –examples of such programmed features are: the PEC junction case, and the simulation of junctions with MoM accelerating techniques–. Additionally, unlike previous algorithms, the above-mentioned algorithm in Section 2 has the mathematical advantage of being elegantly expressed as a compact matrix form which provides valuable insight to the reader on important implementation aspects.

All the meshes used in the simulation examples presented in this paper were generated in the 3D meshing software Gmsh (http://geuz.org/gmsh/) tool, which is distributed under the terms of the GNU General Public License [13]. This free software tool is specially useful for generating surfaces meshes suitable for our SIE generalization, avoiding duplication of unknowns.

The remaining parts of this paper have been structured as follows. In Section 2, we present our algorithm for enforcing current continuity at junction edges. This algorithm relies on the mathematical generalization of SIE formulations for the case of multibody problems which is detailed in the Appendix. Section 3 includes practical simulation results which validate our scheme by comparison with measurements covering different electrical sizes of a grounded monopole antenna for airplane-satellite communication at the IEEE C-band. This practical example includes general dielectric-PEC junctions. Good agreement, in radiation patterns and gains, is observed between MoM and precise measurement results. Finally, Section 4 concludes the manuscript with a summary.

## 2. Algorithm for enforcing current continuity at mesh junctions

A computational technique is described in this section to account for the current continuity at mesh junctions. In our junction method approach, this technique is used with the generalized SIE explained in the Appendix. The current continuity comprises two different current matchings: Kirchhoff's Law relates the current flow inside each region across a junction edge from one separation surface to another, and boundary conditions relate the currents on the outer face of a given surface to those on the inner face of the same surface. In the algorithm presented in this section, boundary conditions are only applied to fictitious RWG bases defined to ensure Kirchhoff's Law at junction edges, while the boundary conditions on the ordinary RWG basis functions defined for the original meshes, passed through to the MoM code, are imposed by the SIE generalization in the Appendix.

The algorithm in this section considers independent meshes for each one of the interfaces that separate any two regions, as it can be seen in the example in Fig. 2. A pair of fictitious triangles, which defines a full RWG basis in the junction method, is formed by two edge triangles of the original independent separation meshes. These original edge triangles, painted dark blue in the example in Fig. 2, have one side located at a geometric edge of each individual mesh, which corresponds to a junction edge of the whole composite geometry.

The method for creation of virtual sets of RWG functions used to ensure Kirchhoff's Current Law –each region in the geometry define a different set as shown in Fig. 3– is explained in 2.1. The virtual sets of fictitious RWG bases are treated in the MoM matrix calculation part, described in the Appendix, as if they belonged to the original surface meshes inputted to the MoM code. Then, in 2.2, we address the method for enforcing boundary conditions on the previously defined fictitious bases.

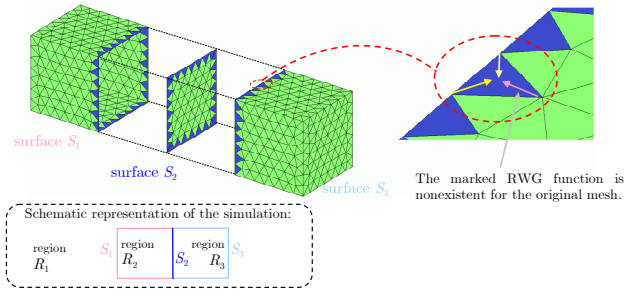

Figure 2. The two touched cuboids in Fig. 1 are modeled using three different separation meshes. The wide gaps depicted among the three separation meshes are solely for representation clarity (in practice all the meshes stay in place as in Fig. 1). The triangles which have only RWG basis functions which do not belong to junction edges are painted green. The triangles in dark blue will be matched to form fictitious ordinary RWG functions for modeling the edge currents (see Fig. 3). Note that the marked RWG basis is not defined for the original mesh, as ordinary RWG bases are only associated to triangle sides which separate two different triangles in a mesh. However, the marked RWG basis will form a fictitious ordinary RWG.

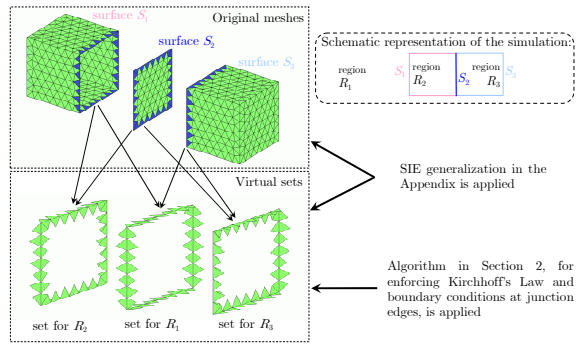

Figure 3. Simulated geometry: (top) original meshes for the separation surfaces in the example of Fig. 1 and Fig. 2; (bottom) sets of virtual RWG functions for each region, formed pairing edge triangles (dark blue triangles) among the original meshes. Each virtual set is only considered for MoM interactions in the region for which it was defined. A full RWG basis is assigned to each pair of adjacent triangles in the virtual sets.

### 2.1. Imposition of Kirchhoff's Law for surface currents at junction edges

Those triangles in a mesh inputted to the code which have a border which coincides with a junction edge are paired together among different separation surface meshes in order to shape what in this paper are called fictitious pairs. The algorithm for defining the sets of fictitious triangle pairs is as follows: 1) first the Euclidean distance between the midpoints $\mathbf{v}_i$ and $\mathbf{v}_j$ of edge sides of two triangles belonging to different original surface meshes is computed; 2) if this Euclidean distance is less than $\delta$, i.e. $|\mathbf{v}_i - \mathbf{v}_j| < \delta$, then a fictitious RWG basis function is defined for these two triangles. The value $\delta > 0$ is introduced to alleviate small alignment errors in the meshes. The procedure is repeated until a fictitious triangle pair sharing the same edge midpoint is found for each surrounding region, as seen in Fig. 4. The edges and triangles that satisfy the aforementioned condition of $|\mathbf{v}_i - \mathbf{v}_j| < \delta$ can be easily found. The meshing software Gmsh [13] was used to generate the interface meshes for this work, and we employed $\delta = \lambda_0/1000$ ($\lambda_0$ is the wavelength in free space).

The virtual sets of fictitious triangle pairs are created to enforce Kirchhoff's Law. Each set defined for the same region and for the same junction edge is treated in the MoM code as if the set consists of a mesh corresponding to an ordinary surface $S_k$, as it is the case with the original surface meshes generated by a meshing software and inputted to the preprocessing part of any MoM code. In order to apply the generalized SIE in the Appendix, a special negative integral index, for instance $R_{in}(k) = -2$ ($R_{in}(k) = -1$ is

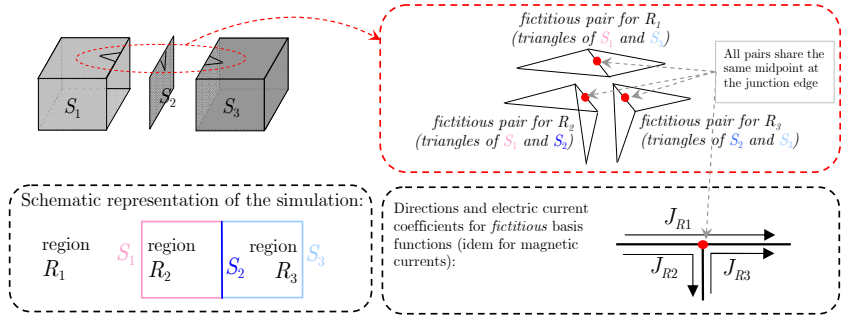

Figure 4. Pairing example which implicates three different triangles which have matching edge sides. A different fictitious pair is defined for each region. A basis function is assigned to each virtual pair. The represented directions are assigned in this example to each virtual basis function in order to avoid changing the sign of each function coefficient when enforcing boundary conditions, which will be explained in Subsection 2.2.

reserved for PEC surfaces), must be assigned to label the internal region of each virtual set (see Appendix for details on labeling). This special index indicates that only the MoM interactions via the outer region $R_{out}(k)$ for which the virtual set was defined must be considered in the MoM code. The code preparation for the special index $R_{in}(k) = -2$ can be easily implemented because it just requires discarding one medium in each MoM interaction, as in the case of PEC surfaces. Once the virtual sets are created, the problem to which the generalized SIE formulation in the MoM code is applied is exemplified in Fig 5.

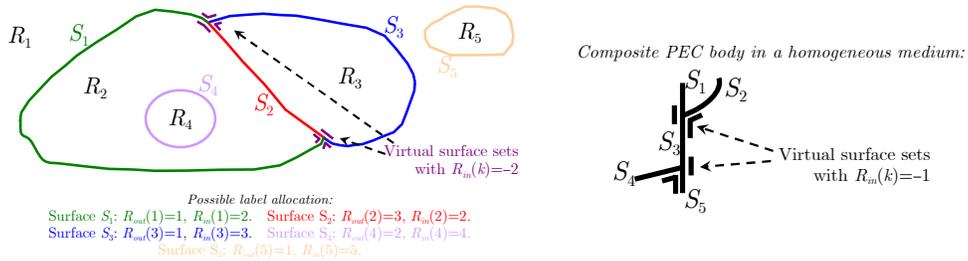

Figure 5. Schematic representation of virtual sets in two examples. The gaps between the virtual sets and the original surfaces are solely for representation clarity. Both the original surfaces and the virtual sets are treated equally in the MoM code. In the special case that the original surfaces which form a junction edge are open PEC surfaces with the same assigned $R_{out}(k)$, which corresponds to a surrounding homogeneous medium, one of the virtual sets must be dismissed for each junction edge so as not to obtain an overdetermined system.

### 2.2. Imposition of boundary conditions at junction edges

If we tried to solve the MoM linear system without enforcing the boundary conditions at junction edges, then the system would have redundancy, and would become unstable. A method for imposing the boundary conditions when virtual sets are employed is described below. This method consists of an unknown-reducing technique which is applied after the existing code calculates the MoM matrix and before the linear system is solved.

For simplicity, without generality loss, we assume that all the regions which converge at any given junction edge are different. If these regions did not have different labels, a version of the explained technique for boundary conditions would still be applicable, but the technique explanation and the computational implementation would become noticeably more complex. Nevertheless, it can be noticed that a real problem involving junction edges comprising repeated regions can always be straightforwardly transformed into an equivalent problem without repeated regions, by simply introducing additional

separation surfaces between regions defined with the same complex permittivities and permeabilities.

The following computational algorithm described for junction edges may seem abstract and difficult to grasp at first. Nonetheless, its underlying simplicity will be more apparent when addressing the example at the end of this subsection. In order to implement our reduction algorithm, the following information must be stored. These data can be generated during the pairing process explained in the preceding subsection:

- A set of two labels $[1, CR(m,n)]$ must be assigned to identify one of the triangles in each fictitious pair, and another pair $[2, CR(m,n)]$ must be assigned to the other triangle. $CR(m,n)$ is the function which gives the domain region common to surfaces $S_m$ (labeled in the code with surface index $m \in \mathbb{N}$) and $S_n$ (labeled with $n \in \mathbb{N}$) to which the two paired triangles $[1, CR(m,n)]$ and $[2, CR(m,n)]$ respectively belong in their original meshes:

$$CR(m,n) = \begin{cases} R_{out}(m) \text{ if } R_{out}(m) = R_{out}(n) \text{ or } R_{out}(m) = R_{in}(n), \\ R_{in}(m) \text{ if } R_{in}(m) = R_{out}(n) \text{ or } R_{in}(m) = R_{in}(n), \\ 0 \text{ otherwise.} \end{cases} \quad (1)$$

  $R_{out}(m) \in \mathbb{N}$ and $R_{in}(m) \in \mathbb{N}$ are, respectively, the labels for the outer and inner regions of surface $S_m$.

- A matrix $\bar{\mathbf{D}}_{region}$ with dimensions $2 \times N_R$, where $N_R$ is the number of regions in the simulation, must be initialized to an all-zeros pattern. Then, whenever a pair of fictitious triangles is formed as explained in 2.1, matrix $\bar{\mathbf{D}}_{region}$ must be updated as follows:

$$\begin{aligned} \bar{\mathbf{D}}_{region}[1, CR(m,n)] &\leftarrow \begin{cases} R_{out}(m) \text{ if } CR(m,n) \neq R_{out}(m), \\ R_{in}(m) \text{ if } CR(m,n) \neq R_{in}(m). \end{cases} \\ \bar{\mathbf{D}}_{region}[2, CR(m,n)] &\leftarrow \begin{cases} R_{out}(n) \text{ if } CR(m,n) \neq R_{out}(n), \\ R_{in}(n) \text{ if } CR(m,n) \neq R_{in}(n). \end{cases} \end{aligned} \quad (2)$$

  In the preceding assignations, $\bar{\mathbf{D}}_{region}[i, CR(m,n)]$, with $i = 1, 2$, represents the element in row $i$ and column $CR(m,n)$ of matrix $\bar{\mathbf{D}}_{region}$. The update process in (2), simply means that the labels of the non-common regions in the original surfaces are stored.

- A vector $\mathbf{V}_{unk}$ must be employed to store sequential labels assigned to identify the unknown index of each created pair: $\mathbf{V}_{unk}[CR(m,n)] \leftarrow 1, 2, 3, \ldots$. An assignation to $\mathbf{V}_{unk}$ is performed whenever a pair of fictitious triangles is formed. The assigned value is increased by one unity in every assignation to $\mathbf{V}_{unk}$.

Finally, it is also required to count the number of fictitious pairs $N_{unk}$ and the number of PEC surfaces $N_{PEC}$ which share the same midpoint. The algorithm for generating a reduction submatrix $\bar{\mathbf{R}}_{midpoint}$ and also for assigning directions to the fictitious RWG bases is detailed in Fig. 6. Matrix $\bar{\mathbf{R}}_{midpoint}$ has dimensions $N_{unk} \times \max\{1, N_{PEC}/2\}$ and, once generated, contains ones and zeros only.

Employing the traditional boundary conditions for the electric and magnetic currents, it is simple to prove that if the basis functions of fictitious triangles belonging to adjacent regions have opposite directions, then the boundary conditions merely imply making those basis function coefficients equal to each other. This assignation of directions to the basis functions is already implemented in the algorithm detailed in Fig. 6. The assignation of directions is shown by the black arrows near the midpoint in the clarifying example in Fig. 7.

**Initialize:** $\bar{\mathbf{R}}_{midpoint} \leftarrow \begin{pmatrix} 0 & 0 & \cdots \\ 0 & 0 & \cdots \\ \vdots & \vdots & \ddots \end{pmatrix}_{N_{unk} \times \max\{1, N_{PEC}/2\}}$, $column \leftarrow 1$.

**If** $N_{PEC} > 0$, **find any** $t$, $r$ **that satisfy:** $\bar{\mathbf{D}}_{region}[t, r] = -1$. % find a PEC triangle
**If** $N_{PEC} = 0$, **find any** $t$, $r$ **that satisfy:** $\bar{\mathbf{D}}_{region}[t, r] > 0$. % find any triangle

**While** $\sum_i \sum_j \bar{\mathbf{R}}_{midpoint}[i, j] < N_{unk}$ **do**

    **Assign RWG direction to the surface currents on triangle with labels** $[t, r]$**: direction** *toward* **the midpoint.**
    **Assign RWG direction to the surface currents on triangle with labels** $[\mathrm{mod}(t,2)+1, r]$**: direction** *from* **the midpoint.**
    **Update the reduction submatrix putting an 1:** $\bar{\mathbf{R}}_{midpoint}[\mathbf{V}_{unk}[r], column] \leftarrow 1$.

    **If** $\bar{\mathbf{D}}_{region}[\mathrm{mod}(t,2)+1, r] \neq -1$, **find any** $t'$, $r'$ **such that:** $\bar{\mathbf{D}}_{region}[t', r'] = r$ **and** $\bar{\mathbf{D}}_{region}[\mathrm{mod}(t,2)+1, r] = r'$.
    % find the triangle which shares the interface with the triangle with labels $[\mathrm{mod}(t,2)+1, r]$.

    **If** $\bar{\mathbf{D}}_{region}[\mathrm{mod}(t,2)+1, r] = -1$, **do**    % a PEC separation was found and a new column has to be filled
        **Find any** $t'$, $r'$ **such that:** $\bar{\mathbf{D}}_{region}[t', r'] = -1$ **and** $\sum_j \bar{\mathbf{R}}_{midpoint}[\mathbf{V}_{unk}[r'], j] = 0$. % find a new PEC triangle not already reduced
        **Update the column counter:** $column \leftarrow column + 1$.
    **end if**
    **Update** $t$, $r$**:** $t \leftarrow t'$, $r \leftarrow r'$.
**end while**

Figure 6. Algorithm for filling a reduction submatrix and for assigning directions to currents.

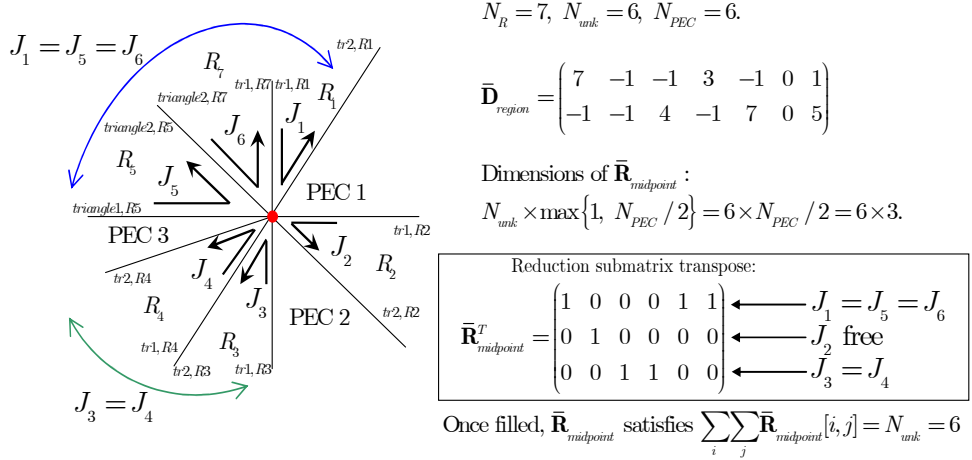

Figure 7. Example of reduction submatrix and direction assignation to RWG bases in a junction edge midpoint. Each straight segment line in the schematic drawing represents a different virtual triangle defined for a region. The assignation triangle1/triangle2 is arbitrary and depends on the order in which the original surface meshes are inspected. The junction edge midpoint is represented by the central red dot. The colored continuous smoothly bent arrows over the drawing represent those unknown current coefficients which must be equal if the boundary conditions are applied using the same directions represented by the black arrows over the drawing for the RWG basis functions.

The global reduction matrix for the full multibody problem can be expressed as a block diagonal matrix which is highly sparse:

$$\bar{\mathbf{R}}_{global} = \begin{pmatrix} \bar{\mathbf{R}}^J_{midpoint1} & \bar{\mathbf{0}} & \bar{\mathbf{0}} & \cdots \\ \bar{\mathbf{0}} & \bar{\mathbf{R}}^M_{midpoint1} & \bar{\mathbf{0}} & \cdots \\ \bar{\mathbf{0}} & \bar{\mathbf{0}} & \bar{\mathbf{R}}^J_{midpoint2} & \cdots \\ \vdots & \vdots & \vdots & \ddots \end{pmatrix}. \quad (3)$$

Each diagonal block in (3) contains a reduction submatrix for each midpoint and for each

type of currents, obtained according to the algorithm in Fig. 6: electric currents $\bar{\mathbf{R}}^J_{midpoint}$ and, if the junction does not involve a PEC region, also magnetic currents $\bar{\mathbf{R}}^M_{midpoint}$. It holds that the algorithm in Fig. 6 is valid for both types of currents: $\bar{\mathbf{R}}^J_{midpoint} = \bar{\mathbf{R}}^M_{midpoint}$.

Once the MoM matrix is calculated and before proceeding with the solution of the multibody linear system $\bar{\mathbf{Z}}_{multi} \cdot \mathbf{I}_{multi} = \mathbf{V}_{multi}$ in (A25), we need to impose the boundary conditions at junction edges, which can be implemented by reducing the number of unknowns as follows:

$$\bar{\mathbf{R}}^T \bar{\mathbf{Z}}_{multi} \bar{\mathbf{R}} \cdot \mathbf{I}'_{multi} = \bar{\mathbf{R}}^T \mathbf{V}_{multi}, \tag{4}$$

where $\mathbf{I}'_{multi}$ denotes the reduced unknown vector and $\bar{\mathbf{R}}$ is a reduction matrix which combines (sums) the columns of MoM matrix $\bar{\mathbf{Z}}_{multi}$ which represent the same unknowns. Similarly, matrix $\bar{\mathbf{R}}^T$ combines the rows of $\bar{\mathbf{Z}}_{multi}$ so that the number of equations in the reduced system is equal to the number of unknowns.

In order to clear up the complete matrix structure in the algorithm, let us assume that the number of unknowns corresponding to all the original surfaces inputted to the code in a multibody simulation is $K$, and that the number of unknowns corresponding to all the virtual sets generated by the code itself is $P$. We include in $K$ and $P$ the magnetic and the electric current coefficients. We denote by $P'$ the number of unknowns in the virtual sets after reduction. Taking into account the explained notation, reduction matrix $\bar{\mathbf{R}}$ in (4) can be finally expressed as a sparse matrix containing four submatrices:

$$\bar{\mathbf{R}} = \begin{pmatrix} \bar{\mathbf{I}}_{K \times K} & \bar{\mathbf{0}}_{K \times P'} \\ \bar{\mathbf{0}}_{P \times K} & \bar{\mathbf{R}}_{global} \end{pmatrix}, \tag{5}$$

where $\bar{\mathbf{I}}_{K \times K}$ is an identity submatrix, and $\bar{\mathbf{0}}_{P \times K}$ and $\bar{\mathbf{0}}_{K \times P'}$ are all-zero submatrices. $\bar{\mathbf{R}}_{global}$ is the block diagonal matrix in (3). After solving the reduced MoM linear system in (4) for $\mathbf{I}'_{multi}$, it is necessary to get the original unknown vector $\mathbf{I}_{multi}$ as follows:

$$\mathbf{I}_{multi} = \bar{\mathbf{R}} \cdot \mathbf{I}'_{multi}. \tag{6}$$

When implementing the reduction in a code, the process can be efficiently accomplished because matrix $\bar{\mathbf{R}}$ is highly sparse. The described matrix algorithm can also be easily implemented in MoM-based parallel and accelerating techniques such as the fast multipole method (FMM) and the multilevel fast multipole algorithm (MLFMA). The reader is referred to [27] for a detailed description of the FMM and the MLFMA algorithms. An overview explanation about using our algorithm together with the MLFMA was published in [28].

## 3. Implementation details and practical simulation results

### 3.1. *Code implementation decisions*

Our implementation of the MoM-SIE method consists of a pure C code with double-precision floating-point calculations. We have employed 7-point Gaussian quadrature rules for numerical integration, together with the analytical extraction methods summarized in Chapter 8 of [27] for the accurate evaluation of the singular integrals. Furthermore, we have applied a diagonal preconditioner for each SIE formulation, as described in 9.6.1 of [27], together with the preconditioning technique in [29]. The MoM system was solved by direct LU decomposition.

Regarding the meshes, we have chosen the "frontal" mesh type in the software Gmsh [13]. This kind of mesh was selected because it provides triangles with good aspect ratios required to obtain accurate results with MoM. The discretization length $\ell_{discr}$, that is the maximum side length for the triangles in the mesh, was set in our simulations to the value $\ell_{discr} = 0.08\lambda_0$, where $\lambda_0$ is the wavelength in free space. This is the same $\ell_{discr}$ value utilized in [30], and it does not introduce a significant error whenever $|k_{medium}/k_0| \leq 3$, with $k_0 = 2\pi/\lambda_0$.

The code was automatically parallelized using the source-to-source compiler Parallware [31, 32]. Parallware is a new parallelizing compiler which uses a novel disruptive technology for the automatic parallelization of C codes. This new technology puts in value recent R&D results in the area of advanced compiler techniques [33, 34]. Thereby the parallelization by hand, which is a time-consuming error-prone task that requires HPC skills by the programmer, is avoided. Parallware automatically extracts the parallelism implicit in the source code of any sequential simulation program written in C. In addition, Parallware automatically generates a parallel-equivalent program written in C and annotated with OpenMP [35] compiler pragmas.

The following results in this section are two representative practical cases where the explained junction algorithm is applied. For an overview of some other simulation examples where the described junction method is also used, the reader is referred to [36].

### 3.2. *Application example: real coaxially-fed antenna for airplane-satellite communication*

The real monopole coaxially-fed antenna for airplane-satellite communication shown in Fig. 8 and designed in [37] was simulated. Even though rotationally symmetric antennas can be analyzed with MoM versions for bodies of revolution (BOR) or using wire-surface basis functions, a complete 3D MoM implementation was employed for our simulations. 3D MoM has many advantages such as the possibility of directional gain optimization by introducing asymmetries. The coaxial dielectric is teflon (PTFE) with relative permittivity $\epsilon_{rel} = 2.1$, and a TEM field $\mathbf{E}_t(\rho) = \frac{V_0}{\rho \ln(b/a)}\hat{\boldsymbol{\rho}}$ (in this equation, $\rho$ is the cylindrical radial distance from the $\hat{\mathbf{z}}$ axis, and $V_0$ stands for the coaxial feed voltage) is imposed in the SIE-MoM method at the bottom of surface $S4$ in Fig. 8. The reflection coefficient $\rho_a$ on surface $S2$ can be obtained applying the matrix pencil method [38] to the MoM surface currents inside the coaxial cable, as described in [39]. Due to infinite reflections at $S2$ and at the short circuited end on surface $S4$ in the model, to calculate the antenna gain $G$ through simulation, an equivalent feed voltage $V_0' = V_0/(1+\rho_a e^{-j2k_{\text{PTFE}}\ell_c})$ ($k_{\text{PTFE}}$ is the wavenumber for PTFE, $\ell_c$ is the simulated cable length) has to be considered for the total feed power, instead of the real voltage $V_0$ defined above for $\mathbf{E}_t$.

Five different SIE formulations have been implemented in our MoM code for the dielectric-air interface in Fig. 8. These five formulations are known in the literature as PMCHWT, CTF, CNF, MNMF and JMCFIE (see Appendix). Perfect electric conductor (PEC) surfaces were analyzed in MoM using the EFIE formulation. The simulation results are summarized in Fig. 9 and Fig. 10. For comparison purposes in the antenna gain, actual measured outcomes and PO (physical optics) results are also included in Fig. 9. The $\epsilon_{rel}$ is not considered in our PO implementation. Hence, the reflection coefficient is only obtained with MoM.

Fig. 9a shows that the measured gain values are in variable agreement with MoM and PO results; however, the average difference in dB is small. The best coincidence in Fig. 9a around the resonance is obtained with the PMCHWT, CTF and JMCFIE formulations. Finally, Fig. 10 exhibits gain results and current magnitudes which evidence the actual importance of correctly imposing Kirchhoffs Law at junction edges, that is, the importance of using the virtual sets in our algorithm for continuity of surface currents.

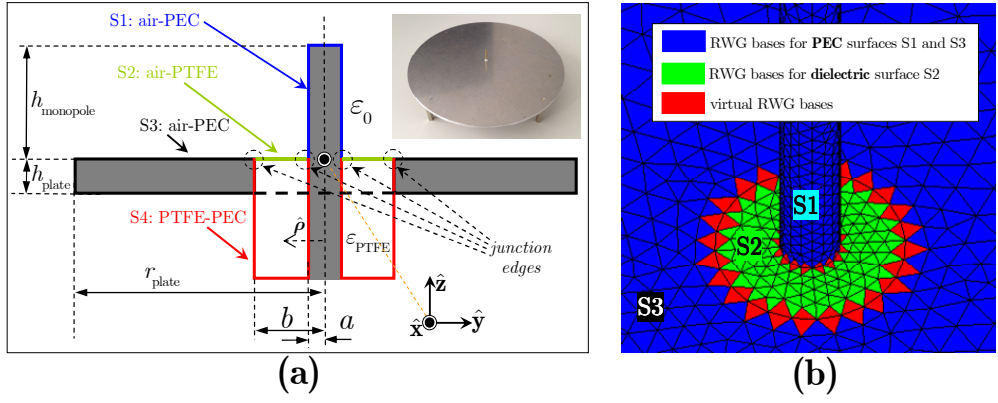

Figure 8. Simulated antenna: (a, top-right) actual antenna picture; (a, rest) schematic –not at scale– of the four surface meshes used to simulate with MoM. $h_{\text{monopole}}$=13.4 mm, $h_{\text{plate}}$=3 mm, $r_{\text{plate}}$=9 cm, $a$=0.64 mm, $b$=2.03 mm; (b) types of RWG bases in different meshes: triangles with only RWG bases for PEC surfaces are in blue, triangles with only RWG bases for dielectric separation surfaces are in green, and triangles with an edge which separates more than two surfaces, over which a virtual RWG basis is defined, are painted red in the figure.

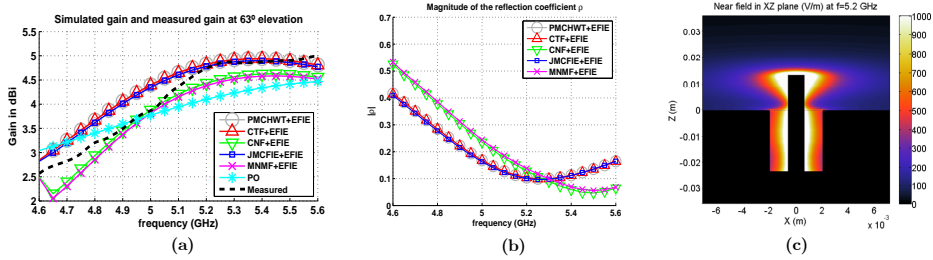

Figure 9. Simulated field results and measurements, and circuit parameters: (a) real measured data, together with simulated PO and MoM antenna gain $G$ in dBi; (b) reflection coefficient; (c) near field with the JMCFIE formulation.

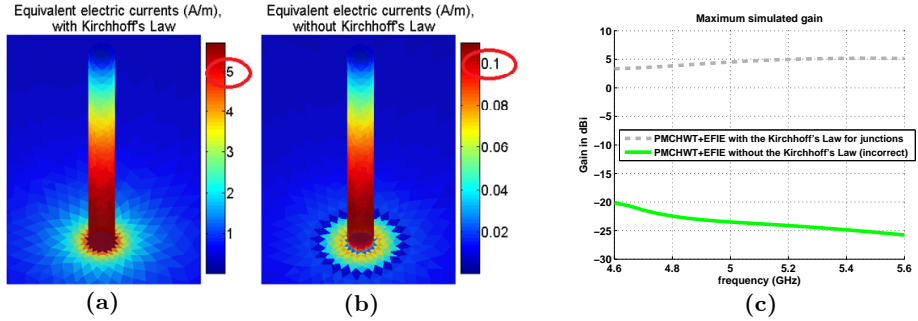

Figure 10. Equivalent-current magnitudes at 5.2 GHz for $V_0$=1 V: (a) *equivalent* surface currents at each triangle centroid, in A/m, with Kirchhoff's Law (the virtual sets of RWG functions are included in the simulation); (b) *equivalent* currents, without Kirchhoff's Law at junction edges (incorrect!); (c) gain vs. frequency with and without current continuity.

### 3.3. *Application example: importance of not duplicating unknowns and enforcing Kirchhoff's Law*

An example is addressed in this subsection to stress the actual importance of two different aspects present in our junction method: avoiding duplicate unknowns and enforcing Kirchhoff's Law. The topic on the importance of Kirchhoff's Law was already partially tackled in the preceding subsection for PEC junctions.

For this example, the incident plane wave comes from vacuum and is defined by the

11

following parameters: amplitude 1 V/m, $f = 60$ GHz, polarization $\hat{\mathbf{x}}$, and incidence direction $-\hat{\mathbf{z}}$. The simulation, summarized in Fig. 11, consists of a non-homogeneous dielectric wedge with 4 spatially varying relative permittivities from $\varepsilon_{r1} = 2$ to $\varepsilon_{r4} = 8$, whereas the following values were used for all the media: $\sigma = 0$, $\mu_r = 1$. The wedge problem was selected because it consists of a representative general diffraction problem on a composite structure. This wedge problem involves 17000 unknowns, counting both electric and magnetic current coefficients. The JMCFIE formulation was selected for the simulations, even though the discussions regarding the relevant aspects are extensible to every SIE formulation. Fig. 11 also includes near-field results obtained with the aforesaid JMCFIE formulation.

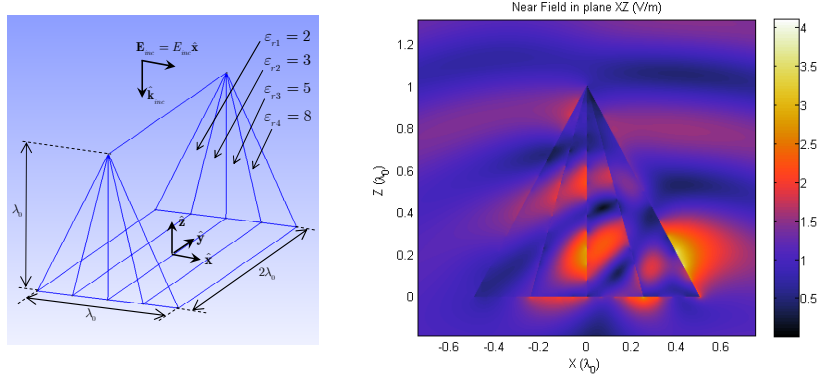

Figure 11. Non-homogeneous dielectric wedge made of 4 dielectric media and modeled with 7 separation surfaces, and total near field obtained in the XZ plane with the JMCFIE formulation. Near-field values inside the wedge are not scaled up.

Far-field results obtained by applying our junction method are compared with FEKO results [40] in Fig. 12. This figure also includes a comparison with the case where separation surfaces are employed but the fictitious RWG functions are not created in the MoM code. In spite of not obtaining an error level as high as in the antenna case, the error peaks present in the case without fictitious bases allow for assessment of the actual importance of correctly enforcing Kirchhoff's Law.

Finally, we must also emphasize in this particular type of simulation the fact that when introducing gaps for approximating without a junction method, or when using a junction method which requires duplicated unknowns, unlike in the method in this paper, the total number of unknowns passes from 17000 to 25000. The increase factor for the number of unknowns is 1.47 with respect to the case where our junction method is employed, resulting in a serious deterioration of the computational performance for large problems in terms of CPU time and RAM memory usage.

## 4. Conclusions

This paper presents a junction method for simulating with accuracy scattering problems where bodies in contact are considered. Compared to previous approaches, in order to incorporate the junction functionality to a MoM code, the implementation of the proposed method does not require deep modifications in existing MoM codes which can deal with multibody problems. The method implementation requires adding, before the MoM matrix calculation, a piece of code for generating sets of virtual triangles. After the MoM matrix calculation, another piece of code must be implemented to perform a reduction in the number of unknowns.

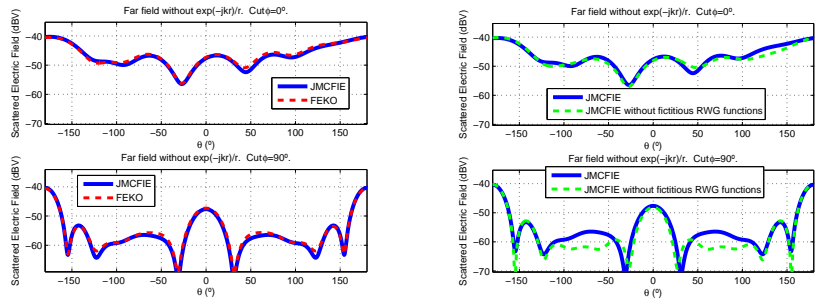

Figure 12. Comparison of JMCFIE results for junctions: (left) FEKO fields are compared with those obtained from our implemented JMCFIE; (right) JMCFIE fields are represented with and without the fictitious RWG bases in our junction method.

An approach for obtaining generalized SIE formulations for multibody problems is also detailed in the Appendix. The resulting system matrix in this SIE generalization approach can be expressed in compact block structure which, unlike other approaches, does not involve a duplication in the number of unknowns, nor does it require including Kirchhoff's Law in the SIE formulation itself. This latter fact allows an easy description and implementation as our SIE generalization does not rely on special extended bases to achieve surface-current continuity.

The method is validated through some representative radiation and scattering examples which include, respectively, a real antenna and a general diffraction case.

## Appendix A. Surface integral equation formulations

Surface integral equations (SIE) are commonly used in computational electromagnetics for obtaining the solution of radiation and scattering problems. For dielectric objects consisting of homogeneous media, SIE are deduced by testing the boundary conditions on the inner and on the outer sides of object surfaces and then combining equations in a proper manner.

In this Appendix, we first explain in Subsection A.1 ordinary SIE formulations for isolated objects surrounded by an unbounded medium. Then, using notation introduced in A.1, Subsection A.2 deals with the generalization of SIE formulations for general

multibody problems.

### A.1. *Surface integral equations for isolated bodies*

Let us consider a homogeneous penetrable scatterer surrounded by a homogeneous unbounded medium. We denote by $R_1$ the region corresponding to the unbounded medium from where an incident wave originates, and denote by $R_2$ the delimited region corresponding to the scattering body. We assume that the incident wave impinges on the separation surface between $R_1$ and $R_2$. For abbreviation, from now on let us associate a subscript $i = 1$ for all the quantities corresponding to $R_1$ (first medium) and a subscript $i = 2$ for all the quantities corresponding to $R_2$ (second medium). Each medium for $i = 1, 2$ is characterized by its constitutive parameters: the complex permittivity $\varepsilon_{c,i} = \varepsilon_{rc,i} \cdot \varepsilon_0$ (which includes the effects of the conductivity $\sigma_i$) and the complex permeability $\mu_{c,i} = \mu_{rc,i} \cdot \mu_0$. $\varepsilon_{rc,i}$ and $\mu_{rc,i}$ are respectively the complex relative permittivity and the complex relative permeability of the medium in region $R_i$. $\varepsilon_0$ and $\mu_0$ are the constitutive parameters of vacuum. A temporal harmonic dependency $\exp(j\omega t)$ is assumed and not included in the formulations.

Making use of Maxwell's equations and vector Green's theorem, and enforcing boundary conditions, in each region $R_i$ the electric field integral equation (EFIE) and the magnetic field integral equation (MFIE) can be formulated, by virtue of Love's field equivalence principle, in two different ways for both EFIE and MFIE [14, 15]. The tangential (T) equations can be written as

$$\begin{aligned} \text{T} - \text{EFIE}_1 \text{ medium 1}: \ \vec{0} &= -\mathbf{E}^{\mathbf{inc}}(\mathbf{r})\big|_{\tan} + L_1 \mathbf{J}(\mathbf{r})\big|_{\tan} - K_1 \mathbf{M}(\mathbf{r})\big|_{\tan} - \tfrac{1}{2}\mathbf{M}(\mathbf{r}) \times \hat{\mathbf{n}}(\mathbf{r}). \\ \text{T} - \text{EFIE}_2 \text{ medium 2}: \ \vec{0} &= L_2 \mathbf{J}(\mathbf{r})\big|_{\tan} - K_2 \mathbf{M}(\mathbf{r})\big|_{\tan} + \tfrac{1}{2}\mathbf{M}(\mathbf{r}) \times \hat{\mathbf{n}}(\mathbf{r}). \end{aligned} \tag{A1}$$

$$\begin{aligned} \text{T} - \text{MFIE}_1 \text{ medium 1}: \ \vec{0} &= -\mathbf{H}^{\mathbf{inc}}(\mathbf{r})\big|_{\tan} + K_1 \mathbf{J}(\mathbf{r})\big|_{\tan} + \tfrac{1}{\eta_1^2} L_1 \mathbf{M}(\mathbf{r})\big|_{\tan} + \tfrac{1}{2}\mathbf{J}(\mathbf{r}) \times \hat{\mathbf{n}}(\mathbf{r}). \\ \text{T} - \text{MFIE}_2 \text{ medium 2}: \ \vec{0} &= K_2 \mathbf{J}(\mathbf{r})\big|_{\tan} + \tfrac{1}{\eta_2^2} L_2 \mathbf{M}(\mathbf{r})\big|_{\tan} - \tfrac{1}{2}\mathbf{J}(\mathbf{r}) \times \hat{\mathbf{n}}(\mathbf{r}). \end{aligned} \tag{A2}$$

Similarly, the normal equations (N) are given by

$$\begin{aligned} \text{N} - \text{EFIE}_1 \text{ medium 1}: \ \vec{0} &= -\hat{\mathbf{n}}(\mathbf{r}) \times \mathbf{E}^{\mathbf{inc}}(\mathbf{r}) + \hat{\mathbf{n}}(\mathbf{r}) \times [L_1 \mathbf{J}(\mathbf{r}) - K_1 \mathbf{M}(\mathbf{r})] - \tfrac{1}{2}\mathbf{M}(\mathbf{r}). \\ \text{N} - \text{EFIE}_2 \text{ medium 2}: \ \vec{0} &= \hat{\mathbf{n}}(\mathbf{r}) \times [L_2 \mathbf{J}(\mathbf{r}) - K_2 \mathbf{M}(\mathbf{r})] + \tfrac{1}{2}\mathbf{M}(\mathbf{r}). \end{aligned} \tag{A3}$$

$$\begin{aligned} \text{N} - \text{MFIE}_1 \text{ medium 1}: \ \vec{0} &= -\hat{\mathbf{n}}(\mathbf{r}) \times \mathbf{H}^{\mathbf{inc}}(\mathbf{r}) + \hat{\mathbf{n}}(\mathbf{r}) \times \left[K_1 \mathbf{J}(\mathbf{r}) + \tfrac{1}{\eta_1^2} L_1 \mathbf{M}(\mathbf{r})\right] + \tfrac{1}{2}\mathbf{J}(\mathbf{r}). \\ \text{N} - \text{MFIE}_2 \text{ medium 2}: \ \vec{0} &= \hat{\mathbf{n}}(\mathbf{r}) \times \left[K_2 \mathbf{J}(\mathbf{r}) + \tfrac{1}{\eta_2^2} L_2 \mathbf{M}(\mathbf{r})\right] - \tfrac{1}{2}\mathbf{J}(\mathbf{r}). \end{aligned} \tag{A4}$$

In the above equations, $\mathbf{J}(\mathbf{r})$ and $\mathbf{M}(\mathbf{r})$ denote the, a-priori unknown, induced equivalent surface currents (electric and magnetic currents respectively) on the interface between $R_1$ and $R_2$. $\mathbf{J}(\mathbf{r})$ and $\mathbf{M}(\mathbf{r})$ are vector functions of an arbitrary surface point $\mathbf{r}$ which is defined approaching the surface from $R_1$. Hence we define $\mathbf{J}(\mathbf{r})$ and $\mathbf{M}(\mathbf{r})$ as surface currents in region $R_1$. (For region $R_2$, the currents are simply $-\mathbf{J}(\mathbf{r})$ and $-\mathbf{M}(\mathbf{r})$ in order to fulfill the boundary conditions.) Also, $\hat{\mathbf{n}}(\mathbf{r})$ is the unit vector normal to the surface and pointing towards exterior region $R_1$. $\eta_i = (\mu_{c,i}/\varepsilon_{c,i})^{1/2}$ is the intrinsic impedance in medium $R_i$. Vectors $\mathbf{E}^{\mathbf{inc}}(\mathbf{r})$ and $\mathbf{H}^{\mathbf{inc}}(\mathbf{r})$ respectively represent the incident electric and magnetic fields at surface point $\mathbf{r}$. The integro-differential operators $L_i$ and $K_i$ in

Table A1. Parameters for obtaining five well-documented surface integral equation formulations.

| formulation | $a_i$ | $b_i$ | $c_i$ | $d_i$ |
|---|---|---|---|---|
| PMCHWT | $\eta_i$ | 0 | 0 | $1/\eta_i$ |
| JMCFIE | 1 | 1 | 1 | 1 |
| CTF | 1 | 0 | 0 | 1 |
| CNF | 0 | 1 | 1 | 0 |
| MNMF | 0 | $\mu_{c,i}/(\mu_{c,1}+\mu_{c,2})$ | $\varepsilon_{c,i}/(\varepsilon_{c,1}+\varepsilon_{c,2})$ | 0 |

(A1)–(A4) are defined as

$$L_i \mathbf{X}(\mathbf{r}) = \int_S \left[ j\omega\mu_{c,i} \mathbf{X}(\mathbf{r}') + \frac{j}{\omega\varepsilon_{c,i}} \nabla \left( \nabla' \cdot \mathbf{X}(\mathbf{r}') \right) \right] G_i(\mathbf{r},\mathbf{r}') \, ds',$$
$$K_i \mathbf{X}(\mathbf{r}) = \oint_S \mathbf{X}(\mathbf{r}') \times \nabla G_i(\mathbf{r},\mathbf{r}') \, ds'. \tag{A5}$$

The symbol $\oint$ is used in the definition of $K_i$ for indicating that the integration is taken as a Cauchy principal value integral. The integration surface $S$ refers to the separation interface between $R_1$ and $R_2$. The term $G_i(\mathbf{r},\mathbf{r}')$ in (A5) is the scalar Green's function in region $R_i$:

$$G_i(\mathbf{r},\mathbf{r}') = \frac{\exp(-jk_i |\mathbf{r}-\mathbf{r}'|)}{4\pi |\mathbf{r}-\mathbf{r}'|}. \tag{A6}$$

In (A5) and (A6), vector $\mathbf{r}'$ stands for a source point on the interface and $\mathbf{r}$ denotes an observation point. In addition, scalar $k_i = \omega(\varepsilon_{c,i}\mu_{c,i})^{1/2}$ is the wavenumber in $R_i$.

A general SIE formulation can be set up based on the different EFIEs and MFIEs in (A1)–(A4). We perform a generic combination of these equations using the same sign conventions as in [15]:

$$\frac{a_1}{\eta_1}(\text{T}-\text{EFIE}_1) + \frac{a_2}{\eta_2}(\text{T}-\text{EFIE}_2) + b_1(\text{N}-\text{MFIE}_1) - b_2(\text{N}-\text{MFIE}_2) = \vec{0},$$
$$-c_1(\text{N}-\text{EFIE}_1) + c_2(\text{N}-\text{EFIE}_2) + d_1\eta_1(\text{T}-\text{MFIE}_1) + d_2\eta_2(\text{T}-\text{MFIE}_2) = \vec{0}. \tag{A7}$$

There are literally infinite values that can be assigned to the complex scalar parameters $a_i, b_i, c_i, d_i$ for $i=1,2$ in order to obtain valid stable formulations. The comparative study included in Section 3 for junctions considers five formulations which are well documented in the case of isolated bodies. The parameters in (A7) which allow to obtain the five considered formulations can be consulted in Table A1. These formulations are known as Poggio-Miller-Chang-Harrington-Wu-Tsai (PMCHWT) [16–19], combined tangential formulation (CTF) [14, 19], combined normal formulation (CNF) [14, 19], modified normal Müller formulation (MNMF) [20], and electric and magnetic current combined-field integral equation (JMCFIE) [9, 21–24]. Further formulation collections which incorporate other stable formulations can be consulted in [15, 19, 25, 26].

To solve a SIE formulation in the form (A7), the unknown current densities $\mathbf{J}(\mathbf{r})$ and $\mathbf{M}(\mathbf{r})$ are approximated in terms of linear combinations of known vector basis functions $\mathbf{f}_n$, $n=1,\ldots,N$, as

$$\mathbf{J}(\mathbf{r}) = \sum_{n=1}^{N} J_n \mathbf{f}_n, \quad \mathbf{M}(\mathbf{r}) = \sum_{n=1}^{N} M_n \mathbf{f}_n, \tag{A8}$$

where $J_n$ and $M_n$ are the unknown complex coefficients in the expansions in (A8). If the RWG basis functions are chosen, then each $\mathbf{f}_n$ is assigned to a triangle side in the mesh, and each $\mathbf{f}_n$ is only defined over the two triangles adjacent to the assigned side [6].

Applying the Galerkin testing procedure, each testing function assigned to a side has the same vector expression as the corresponding basis function. For simplicity, the testing operation, with a testing function $\mathbf{f}_m$, of a generic vector function $\mathbf{v}(\mathbf{r})$ will be denoted as $\langle \mathbf{f}_m, \mathbf{v}(\mathbf{r}) \rangle = \int_{S_m} \mathbf{f}_m \cdot \mathbf{v}(\mathbf{r}) ds$, where the dot operator inside the integral represents a scalar product, and $S_m$ is the integration area over which $\mathbf{f}_m$ is defined.

Substituting relations (A8) into (A7) and performing MoM testing with functions $\mathbf{f}_m$, $m = 1, \ldots, N$, results in a $2N \times 2N$ system of liner equations:

$$\bar{\mathbf{Z}} \cdot \mathbf{I} = \mathbf{V}. \tag{A9}$$

The MoM matrix $\bar{\mathbf{Z}}$ has the form

$$\bar{\mathbf{Z}} = \begin{bmatrix} \bar{\mathbf{Z}}^{J,(T-EFIE,N-MFIE)} & \bar{\mathbf{Z}}^{M,(T-EFIE,N-MFIE)} \\ \bar{\mathbf{Z}}^{J,(T-MFIE,N-EFIE)} & \bar{\mathbf{Z}}^{M,(T-MFIE,N-EFIE)} \end{bmatrix}, \tag{A10}$$

with the entries of the four $N \times N$ submatrices given by the following expressions for $m = 1, \ldots, N$ and $n = 1, \ldots, N$:

$$Z_{m,n}^{J,(T-EFIE,N-MFIE)} = \left\langle \mathbf{f}_m, \left( \frac{a_1}{\eta_1} L_1 + \frac{a_2}{\eta_2} L_2 \right) \mathbf{f}_n \right\rangle + \langle \mathbf{f}_m, \hat{\mathbf{n}}_m \times (b_1 K_1 - b_2 K_2) \mathbf{f}_n \rangle + \frac{b_1 + b_2}{2} \langle \mathbf{f}_m, \mathbf{f}_n \rangle, \tag{A11}$$

$$Z_{m,n}^{M,(T-EFIE,N-MFIE)} =$$
$$-\left\langle \mathbf{f}_m, \left( \frac{a_1}{\eta_1} K_1 + \frac{a_2}{\eta_2} K_2 \right) \mathbf{f}_n \right\rangle + \frac{1}{2} \left( \frac{a_1}{\eta_1} - \frac{a_2}{\eta_2} \right) \langle \mathbf{f}_m, \hat{\mathbf{n}}_m \times \mathbf{f}_n \rangle + \left\langle \mathbf{f}_m, \hat{\mathbf{n}}_m \times \left( \frac{b_1 L_1}{\eta_1^2} - \frac{b_2 L_2}{\eta_2^2} \right) \mathbf{f}_n \right\rangle, \tag{A12}$$

$$Z_{m,n}^{J,(T-MFIE,N-EFIE)} = \langle \mathbf{f}_m, \hat{\mathbf{n}}_m \times (-c_1 L_1 + c_2 L_2) \mathbf{f}_n \rangle + \langle \mathbf{f}_m, (d_1 \eta_1 K_1 + d_2 \eta_2 K_2) \mathbf{f}_n \rangle - \frac{d_1 \eta_1 - d_2 \eta_2}{2} \langle \mathbf{f}_m, \hat{\mathbf{n}}_m \times \mathbf{f}_n \rangle, \tag{A13}$$

$$Z_{m,n}^{M,(T-MFIE,N-EFIE)} = \langle \mathbf{f}_m, \hat{\mathbf{n}}_m \times (c_1 K_1 - c_2 K_2) \mathbf{f}_n \rangle + \frac{c_1 + c_2}{2} \langle \mathbf{f}_m, \mathbf{f}_n \rangle + \left\langle \mathbf{f}_m, \left( \frac{d_1 L_1}{\eta_1} + \frac{d_2 L_2}{\eta_2} \right) \mathbf{f}_n \right\rangle. \tag{A14}$$

Also, vector $\mathbf{I}$ containing $2N$ unknown coefficients is given by

$$\mathbf{I} = (J_1, J_2, \ldots J_N, M_1, M_2, \ldots M_N)^T, \tag{A15}$$

and the excitation vector $\mathbf{V}$ of the linear system is

$$\mathbf{V} = \begin{pmatrix} \mathbf{V}^{(T-EFIE,N-MFIE)} \\ \mathbf{V}^{(T-MFIE,N-EFIE)} \end{pmatrix} =$$
$$(V_1^{(T-EFIE,N-MFIE)}, \ldots V_N^{(T-EFIE,N-MFIE)}, V_1^{(T-MFIE,N-EFIE)}, \ldots V_N^{(T-MFIE,N-EFIE)})^T \tag{A16}$$

with excitation coefficients

$$V_m^{(T-EFIE,N-MFIE)} = \frac{a_1}{\eta_1} \langle \mathbf{f}_m, \mathbf{E}^{\mathbf{inc}}(\mathbf{r}) \rangle + b_1 \langle \mathbf{f}_m, \hat{\mathbf{n}}_m \times \mathbf{H}^{\mathbf{inc}}(\mathbf{r}) \rangle \quad \text{for} \quad m = 1, \ldots, N,$$
$$V_m^{(T-MFIE,N-EFIE)} = -c_1 \langle \mathbf{f}_m, \hat{\mathbf{n}}_m \times \mathbf{E}^{\mathbf{inc}}(\mathbf{r}) \rangle + d_1 \eta_1 \langle \mathbf{f}_m, \mathbf{H}^{\mathbf{inc}}(\mathbf{r}) \rangle \quad \text{for} \quad m = 1, \ldots, N. \tag{A17}$$

## A.2. *Generalization of surface integral equations for multibody problems*

In this subsection, the general scheme which was previously described for implementing any SIE formulation is generalized for the case of a multibody problem. This kind of problems may involve bodies in contact, disjoint bodies, and completely embedded bodies inside other bodies.

In the case of disjoint or embedded bodies, the generalized SIE presented here is valid for all the triangles, i.e. for all the basis functions, in the meshes which discretize the geometries. If the simulation involves junctions, then the generalized SIE is valid for all the triangles except for those with a side coincident with a junction edge. For considering the effects of these edge triangles, the generalized SIE must be complemented with the techniques described in Section 2.

In order to generalize SIE formulations for a multibody problem, we label each surface $S_k$ with an integer index $k > 0$ and each region $R_p$ with an integer index $p > 0$. Additionally, two more indices are assigned to each surface $S_k$: an outer-region index $R_{out}(k)$ and an inner-region index $R_{in}(k)$. The allocation of outer and inner regions to a particular surface can be done in a random manner. Nevertheless, we follow the rule in the SIE that normal vectors to surface $S_k$ must point toward the region whose label index has been assigned to $R_{out}(k)$, and accordingly the equivalent surface currents which the MoM code will determine correspond to outer region $R_{out}(k)$. As an example, let us consider the schematic domains and the labels over the picture in Fig. A1. In this example in Fig. A1, the allocation exhibited below the picture is not the only one that is acceptable.

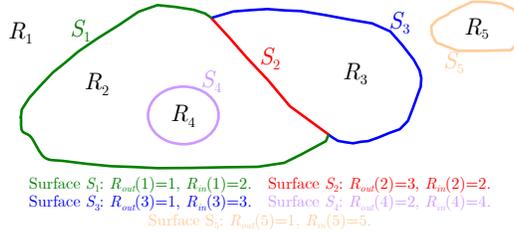

Surface $S_1$: $R_{out}(1)=1$, $R_{in}(1)=2$. Surface $S_2$: $R_{out}(2)=3$, $R_{in}(2)=2$.
Surface $S_3$: $R_{out}(3)=1$, $R_{in}(3)=3$. Surface $S_4$: $R_{out}(4)=2$, $R_{in}(4)=4$.
Surface $S_5$: $R_{out}(5)=1$, $R_{in}(5)=5$.

Figure A1. Schematic representation of five separation surfaces and five volumetric regions. One of many valid allocations for outer and inner regions to each surface is shown below the picture.

The tangential (T) and normal (N) equations, analogous to (A1)–(A4), for calculating the MoM interactions when the observation point $\mathbf{r} \in S_k$ in a problem where multiple surfaces are involved can be written in terms of observation surface labels $k$, source surface labels $k'$, and region labels $R_{out}(k)$ and $R_{in}(k)$:

$$\begin{aligned}
\text{T}-\text{EFIE}_{k,R_{out}(k)}: \ \vec{0} &= -F_{inc}[R_{out}(k)] \cdot \mathbf{E^{inc}}(\mathbf{r})\big|_{\tan} - \tfrac{1}{2}\mathbf{M}_k(\mathbf{r}) \times \hat{\mathbf{n}}_k(\mathbf{r}) + \\
&\quad \sum_{k'} F_{sign,R_{out}(k)}(k') \cdot \left( L_{R_{out}(k)} \mathbf{J}_{k'}(\mathbf{r})\big|_{\tan} - K_{R_{out}(k)} \mathbf{M}_{k'}(\mathbf{r})\big|_{\tan} \right), \\
\text{T}-\text{EFIE}_{k,R_{in}(k)}: \ \vec{0} &= -F_{inc}[R_{in}(k)] \cdot \mathbf{E^{inc}}(\mathbf{r})\big|_{\tan} + \tfrac{1}{2}\mathbf{M}_k(\mathbf{r}) \times \hat{\mathbf{n}}_k(\mathbf{r}) + \\
&\quad \sum_{k'} F_{sign,R_{in}(k)}(k') \cdot \left( L_{R_{in}(k)} \mathbf{J}_{k'}(\mathbf{r})\big|_{\tan} - K_{R_{in}(k)} \mathbf{M}_{k'}(\mathbf{r})\big|_{\tan} \right).
\end{aligned} \quad (A18)$$

$$\text{T}-\text{MFIE}_{k,R_{out}(k)}\colon\quad \vec{0} = -F_{inc}[R_{out}(k)] \cdot \mathbf{H^{inc}}(\mathbf{r})\big|_{\tan} + \tfrac{1}{2}\mathbf{J}_k(\mathbf{r}) \times \hat{\mathbf{n}}_k(\mathbf{r}) +$$
$$\sum_{k'} F_{sign,R_{out}(k)}(k') \cdot \left( K_{R_{out}(k)} \mathbf{J}_{k'}(\mathbf{r})\big|_{\tan} + \tfrac{L_{R_{out}(k)}}{\eta_{R_{out}(k)}^2} \mathbf{M}_{k'}(\mathbf{r})\big|_{\tan} \right),$$

$$\text{T}-\text{MFIE}_{k,R_{in}(k)}\colon\quad \vec{0} = -F_{inc}[R_{in}(k)] \cdot \mathbf{H^{inc}}(\mathbf{r})\big|_{\tan} - \tfrac{1}{2}\mathbf{J}_k(\mathbf{r}) \times \hat{\mathbf{n}}_k(\mathbf{r}) +$$
$$\sum_{k'} F_{sign,R_{in}(k)}(k') \cdot \left( K_{R_{in}(k)} \mathbf{J}_{k'}(\mathbf{r})\big|_{\tan} + \tfrac{L_{R_{in}(k)}}{\eta_{R_{in}(k)}^2} \mathbf{M}_{k'}(\mathbf{r})\big|_{\tan} \right).$$
(A19)

$$\text{N}-\text{EFIE}_{k,R_{out}(k)}\colon\quad \vec{0} = -F_{inc}[R_{out}(k)] \cdot \left( \hat{\mathbf{n}}_k(\mathbf{r}) \times \mathbf{E^{inc}}(\mathbf{r})\big|_{\tan} \right) - \tfrac{1}{2}\mathbf{M}_k(\mathbf{r}) +$$
$$\sum_{k'} F_{sign,R_{out}(k)}(k') \cdot \hat{\mathbf{n}}_k(\mathbf{r}) \times \left( L_{R_{out}(k)} \mathbf{J}_{k'}(\mathbf{r})\big|_{\tan} - K_{R_{out}(k)} \mathbf{M}_{k'}(\mathbf{r})\big|_{\tan} \right),$$

$$\text{N}-\text{EFIE}_{k,R_{in}(k)}\colon\quad \vec{0} = -F_{inc}[R_{in}(k)] \cdot \left( \hat{\mathbf{n}}_k(\mathbf{r}) \times \mathbf{E^{inc}}(\mathbf{r})\big|_{\tan} \right) + \tfrac{1}{2}\mathbf{M}_k(\mathbf{r}) +$$
$$\sum_{k'} F_{sign,R_{in}(k)}(k') \cdot \hat{\mathbf{n}}_k(\mathbf{r}) \times \left( L_{R_{in}(k)} \mathbf{J}_{k'}(\mathbf{r})\big|_{\tan} - K_{R_{in}(k)} \mathbf{M}_{k'}(\mathbf{r})\big|_{\tan} \right).$$
(A20)

$$\text{N}-\text{MFIE}_{k,R_{out}(k)}\colon\quad \vec{0} = -F_{inc}[R_{out}(k)] \cdot \left( \hat{\mathbf{n}}_k(\mathbf{r}) \times \mathbf{H^{inc}}(\mathbf{r})\big|_{\tan} \right) + \tfrac{1}{2}\mathbf{J}_k(\mathbf{r}) +$$
$$\sum_{k'} F_{sign,R_{out}(k)}(k') \cdot \hat{\mathbf{n}}_k(\mathbf{r}) \times \left( K_{R_{out}(k)} \mathbf{J}_{k'}(\mathbf{r})\big|_{\tan} + \tfrac{L_{R_{out}(k)}}{\eta_{R_{out}(k)}^2} \mathbf{M}_{k'}(\mathbf{r})\big|_{\tan} \right),$$

$$\text{N}-\text{MFIE}_{k,R_{in}(k)}\colon\quad \vec{0} = -F_{inc}[R_{in}(k)] \cdot \left( \hat{\mathbf{n}}_k(\mathbf{r}) \times \mathbf{H^{inc}}(\mathbf{r})\big|_{\tan} \right) - \tfrac{1}{2}\mathbf{J}_k(\mathbf{r}) +$$
$$\sum_{k'} F_{sign,R_{in}(k)}(k') \cdot \hat{\mathbf{n}}_k(\mathbf{r}) \times \left( K_{R_{in}(k)} \mathbf{J}_{k'}(\mathbf{r})\big|_{\tan} + \tfrac{L_{R_{in}(k)}}{\eta_{R_{in}(k)}^2} \mathbf{M}_{k'}(\mathbf{r})\big|_{\tan} \right).$$
(A21)

Function $F_{inc}$ above is simply defined as

$$F_{inc}(p) = \begin{cases} 1 & \text{if } p = 1, \\ 0 & \text{otherwise.} \end{cases}$$
(A22)

In the previous expression, we have assumed that Region 1 is the unbounded medium where the incident wave comes from, hence $p = 1$ appears in the first case of the definition of $F_{inc}$. Also, functions $F_{sign,R_{out}(k)}$ and $F_{sign,R_{in}(k)}$ are

$$F_{sign,R_{out}(k)}(k') = \begin{cases} 1 & \text{if } R_{out}(k) = R_{out}(k'), \\ -1 & \text{if } R_{out}(k) = R_{in}(k'), \\ 0 & \text{otherwise,} \end{cases} \qquad F_{sign,R_{in}(k)}(k') = \begin{cases} 1 & \text{if } R_{in}(k) = R_{in}(k'), \\ -1 & \text{if } R_{in}(k) = R_{out}(k'), \\ 0 & \text{otherwise.} \end{cases}$$
(A23)

Functions $F_{sign,R_{out}(k)}$ and $F_{sign,R_{in}(k)}$ are introduced in the above equations (A18)–(A21) to accomplish the following two purposes: *i)* including only surfaces which share a common region with the observation surface $S_k$ and *ii)* changing the sign to the source currents, in order to obtain an equivalent problem as if the currents which the MoM method determines on all the surfaces which share a common region had been defined for the same relative face of each surface, that is, all the currents inside or outside the common region which the surfaces delimit.

The general SIE formulation for the observation surface $S_k$ can be stated as

$$\frac{a_{R_{out}(k)}}{\eta_{R_{out}(k)}} \left[\text{T} - \text{EFIE}_{k,R_{out}(k)}\right] + \frac{a_{R_{in}(k)}}{\eta_{R_{in}(k)}} \left[\text{T} - \text{EFIE}_{k,R_{in}(k)}\right] +$$
$$b_{R_{out}(k)} \left[\text{N} - \text{MFIE}_{k,R_{out}(k)}\right] - b_{R_{in}(k)} \left[\text{N} - \text{MFIE}_{k,R_{in}(k)}\right] = \vec{0}, \tag{A24}$$

$$-c_{R_{out}(k)} \left[\text{N} - \text{EFIE}_{k,R_{out}(k)}\right] + c_{R_{in}(k)} \left[\text{N} - \text{EFIE}_{k,R_{in}(k)}\right] +$$
$$d_{R_{out}(k)}\eta_{R_{out}(k)} \left[\text{T} - \text{MFIE}_{k,R_{out}(k)}\right] + d_{R_{in}(k)}\eta_{R_{in}(k)} \left[\text{T} - \text{MFIE}_{k,R_{in}(k)}\right] = \vec{0}.$$

The weighting coefficients $a_i, b_i, c_i, d_i$ are analogous to those in Table A1, but now replacing there index $i = 1$ with $i = R_{out}(k)$ and index $i = 2$ with $i = R_{in}(k)$.

The linear system of equations for the case of a multibody problem,

$$\bar{\mathbf{Z}}_{multi} \cdot \mathbf{I}_{multi} = \mathbf{V}_{multi}, \tag{A25}$$

can be formulated introducing in (A24), for each separation surface $S_{k'}$, expansions of basis functions $\mathbf{f}_n^{(k')}$, $n = 1, \ldots, N_{k'}$, analogous to those in (A8), and then, on the grounds of the Galerkin procedure, performing MoM testing with a set of functions $\mathbf{f}_m^{(k)}$, $m = 1, \ldots, N_k$ at each observation surface $S_k$. The resulting matrix $\bar{\mathbf{Z}}_{multi}$ can be expressed in terms of submatrices $\bar{\mathbf{Z}}_{(k,k')}$ for each pair of observation surface $S_k$ and source surface $S_{k'}$:

$$\bar{\mathbf{Z}}_{multi} = \begin{bmatrix} \bar{\mathbf{Z}}_{(1,1)} & \bar{\mathbf{Z}}_{(1,2)} & \cdots \\ \bar{\mathbf{Z}}_{(2,1)} & \bar{\mathbf{Z}}_{(2,2)} & \cdots \\ \vdots & \vdots & \ddots \end{bmatrix}. \tag{A26}$$

If $k = k'$, i.e. the observation surface and source surface are the same, then the elements of the corresponding submatrix $\bar{\mathbf{Z}}_{(k,k)}$ include interactions via two different media and the expression for $\bar{\mathbf{Z}}_{(k,k)}$ is the same as in (A10)–(A14). This means that $\bar{\mathbf{Z}}_{(k,k)}$ has the same form as the full MoM matrix in the case of an isolated body, but now substituting $i = R_{out}(k)$ for $i = 1$ and $i = R_{in}(k)$ for $i = 2$. If $k \neq k'$, in order to express the elements of $\bar{\mathbf{Z}}_{(k,k')}$ in compact form, we introduce function $CR(k, k')$ which simply gives the domain region common to surfaces $S_k$ and $S_{k'}$:

$$CR(k, k') = \begin{cases} R_{out}(k) & \text{if } R_{out}(k) = R_{out}(k') \text{ or } R_{out}(k) = R_{in}(k'), \\ R_{in}(k) & \text{if } R_{in}(k) = R_{out}(k') \text{ or } R_{in}(k) = R_{in}(k'), \\ 0 & \text{otherwise.} \end{cases} \tag{A27}$$

Furthermore, an additional function $F_{intra,k}[CR(k, k')]$ is required to compactly model the signs in (A24):

$$F_{intra,k}[CR(k, k')] = \begin{cases} 1 & \text{if } R_{out}(k) = CR(k, k'), \\ -1 & \text{if } R_{in}(k) = CR(k, k'), \\ 0 & \text{otherwise.} \end{cases} \tag{A28}$$

Finally, each submatrix in (A26) for $k \neq k'$ can be written as a $2N_k \times 2N_{k'}$ matrix:

$$\bar{\mathbf{Z}}_{(k,k'),\, CR(k,k')\neq 0} = \begin{bmatrix} \bar{\mathbf{Z}}^{J,(T-EFIE,N-MFIE),(k,k')} & \bar{\mathbf{Z}}^{M,(T-EFIE,N-MFIE),(k,k')} \\ \bar{\mathbf{Z}}^{J,(T-MFIE,N-EFIE),(k,k')} & \bar{\mathbf{Z}}^{M,(T-MFIE,N-EFIE),(k,k')} \end{bmatrix}. \tag{A29}$$

The entries of the four $N_k \times N_{k'}$ preceding submatrices in submatrix $\bar{\mathbf{Z}}_{(k,k')}$ are given by the following expressions, involving the functions previously introduced in (A23), (A27) and (A28), for $m = 1, \ldots, N_k$ and $n = 1, \ldots, N_{k'}$:

$$Z_{m,n}^{J,(T-EFIE,N-MFIE),(k,k')} = F_{sign,CR(k,k')}(k') \left\{ \frac{a_{CR(k,k')}}{\eta_{CR(k,k')}} \left\langle \mathbf{f}_m^{(k)}, L_{CR(k,k')} \mathbf{f}_n^{(k')} \right\rangle + \right.$$
$$\left. F_{intra,k}[CR(k,k')] b_{CR(k,k')} \left\langle \mathbf{f}_m^{(k)}, \hat{\mathbf{n}}_m^{(k)} \times K_{CR(k,k')} \mathbf{f}_n^{(k')} \right\rangle \right\}, \quad \text{(A30)}$$

$$Z_{m,n}^{M,(T-EFIE,N-MFIE),(k,k')} = F_{sign,CR(k,k')}(k') \left\{ \frac{-a_{CR(k,k')}}{\eta_{CR(k,k')}} \left\langle \mathbf{f}_m^{(k)}, K_{CR(k,k')} \mathbf{f}_n^{(k')} \right\rangle + \right.$$
$$\left. F_{intra,k}[CR(k,k')] \frac{b_{CR(k,k')}}{\eta_{CR(k,k')}^2} \left\langle \mathbf{f}_m^{(k)}, \hat{\mathbf{n}}_m^{(k)} \times L_{CR(k,k')} \mathbf{f}_n^{(k')} \right\rangle \right\}, \quad \text{(A31)}$$

$$Z_{m,n}^{J,(T-MFIE,N-EFIE),(k,k')} = F_{sign,CR(k,k')}(k') \left\{ -F_{intra,k}[CR(k,k')] c_{CR(k,k')} \left\langle \mathbf{f}_m^{(k)}, \hat{\mathbf{n}}_m^{(k)} \times L_{CR(k,k')} \mathbf{f}_n^{(k')} \right\rangle + \right.$$
$$\left. d_{CR(k,k')} \eta_{CR(k,k')} \left\langle \mathbf{f}_m^{(k)}, K_{CR(k,k')} \mathbf{f}_n^{(k')} \right\rangle \right\}, \quad \text{(A32)}$$

$$Z_{m,n}^{M,(T-MFIE,N-EFIE),(k,k')} = F_{sign,CR(k,k')}(k') \left\{ -F_{intra,k}[CR(k,k')] c_{CR(k,k')} \left\langle \mathbf{f}_m^{(k)}, \hat{\mathbf{n}}_m^{(k)} \times K_{CR(k,k')} \mathbf{f}_n^{(k')} \right\rangle + \right.$$
$$\left. \frac{d_{CR(k,k')}}{\eta_{CR(k,k')}} \left\langle \mathbf{f}_m^{(k)}, L_{CR(k,k')} \mathbf{f}_n^{(k')} \right\rangle \right\}. \quad \text{(A33)}$$

The excitation vector $\mathbf{V}_{multi}$ of the linear system for a multibody problem can be expressed as an ordered concatenation of excitation subvectors for each different surface:

$$\mathbf{V}_{multi} = \begin{pmatrix} \mathbf{V}_{(1)} \\ \mathbf{V}_{(2)} \\ \vdots \end{pmatrix}, \quad \text{(A34)}$$

where each subvector $\mathbf{V}_{(k)}$ for a surface $S_k$ has the following format:

$$\mathbf{V}_{(k)} = \begin{pmatrix} \mathbf{V}^{(T-EFIE,N-MFIE),(k)} \\ \mathbf{V}^{(T-MFIE,N-EFIE),(k)} \end{pmatrix} =$$
$$(V_1^{(T-EFIE,N-MFIE),(k)}, \ldots V_{N_k}^{(T-EFIE,N-MFIE),(k)}, V_1^{(T-MFIE,N-EFIE),(k)}, \ldots V_{N_k}^{(T-MFIE,N-EFIE),(k)})^T. \quad \text{(A35)}$$

Unlike the entries of the submatrices, the excitation coefficients in the preceding expression do not depend on functions $F_{intra,k}[CR(k,k')]$ and $F_{sign,CR(k,k')}(k')$. These excitation coefficients are, for $m = 1, \ldots, N_k$,

$$V_m^{(T-EFIE,N-MFIE),(k)} = F_{inc}[R_{out}(k)] \left\{ \frac{a_{R_{out}(k)}}{\eta_{R_{out}(k)}} \left\langle \mathbf{f}_m^{(k)}, \mathbf{E}^{\mathbf{inc}}(\mathbf{r}) \right\rangle + b_{R_{out}(k)} \left\langle \mathbf{f}_m^{(k)}, \hat{\mathbf{n}}_m^{(k)} \times \mathbf{H}^{\mathbf{inc}}(\mathbf{r}) \right\rangle \right\} +$$
$$F_{inc}[R_{in}(k)] \left\{ \frac{a_{R_{in}(k)}}{\eta_{R_{in}(k)}} \left\langle \mathbf{f}_m^{(k)}, \mathbf{E}^{\mathbf{inc}}(\mathbf{r}) \right\rangle - b_{R_{in}(k)} \left\langle \mathbf{f}_m^{(k)}, \hat{\mathbf{n}}_m^{(k)} \times \mathbf{H}^{\mathbf{inc}}(\mathbf{r}) \right\rangle \right\}.$$

$$V_m^{(T-MFIE,N-EFIE),(k)} = F_{inc}[R_{out}(k)] \left\{ -c_{R_{out}(k)} \left\langle \mathbf{f}_m^{(k)}, \hat{\mathbf{n}}_m^{(k)} \times \mathbf{E}^{\mathbf{inc}}(\mathbf{r}) \right\rangle + d_{R_{out}(k)} \eta_{R_{out}(k)} \left\langle \mathbf{f}_m^{(k)}, \mathbf{H}^{\mathbf{inc}}(\mathbf{r}) \right\rangle \right\} +$$
$$F_{inc}[R_{in}(k)] \left\{ c_{R_{in}(k)} \left\langle \mathbf{f}_m^{(k)}, \hat{\mathbf{n}}_m^{(k)} \times \mathbf{E}^{\mathbf{inc}}(\mathbf{r}) \right\rangle + d_{R_{in}(k)} \eta_{R_{in}(k)} \left\langle \mathbf{f}_m^{(k)}, \mathbf{H}^{\mathbf{inc}}(\mathbf{r}) \right\rangle \right\}.$$
$$\text{(A36)}$$

Vector $\mathbf{I}_{multi}$ containing the unknown coefficients in a multibody problem can also be stated as a concatenation where each subvector associated to each surface $S_k$ has the following appearance:

$$\mathbf{I}_{(k)} = (J_1^{(k)}, J_2^{(k)}, \ldots J_{N_k}^{(k)}, M_1^{(k)}, M_2^{(k)}, \ldots M_{N_k}^{(k)})^T, \tag{A37}$$

where $J_n^{(k)}$ and $M_n^{(k)}$ are the unknown complex coefficients in the expansions of the basis functions on $S_k$.

### A.3. *Generalization of surface integral equations for multibody problems involving open and closed perfect-electric-conductor bodies*

If one or more of the surfaces in a multibody problem delimit an ideal perfect electric conductor (PEC) object, then a special treatment of the whole problem is required. Without loss of generality, we assume that the index in $R_{out}(k)$ always corresponds to the non-PEC exterior domain associated to a surface $S_k$ which separates a metallic PEC object from another outer medium. In our particular implementation, we merely assign a special interior index $R_{in}(k) = -1$ if $S_k$ delimits a PEC object. It is necessary to rewrite function $F_{sign, R_{in}(k)}$ in (A23) to consider negative indices:

$$F_{sign, R_{in}(k)}(k') = \begin{cases} 1 \text{ if } R_{in}(k) = R_{in}(k') \text{ and } R_{in}(k) > 0, \\ -1 \text{ if } R_{in}(k) = R_{out}(k'), \\ 0 \text{ otherwise.} \end{cases} \tag{A38}$$

First we will assume that a PEC object is delimited in a simulation by a *closed* surface $S_k$ (we call a separation surface closed if it delimits an object with a non-zero volume). The basis and testing functions associated to magnetic currents on a closed $S_k$ delimiting a PEC object must be removed from the formulation, because the boundary conditions entail that $\mathbf{M}_k(\mathbf{r})$ vanishes on PEC surfaces. This means that we must get rid of the coefficients $M_n^{(k)}$ in (A37).

Let us now assume that a PEC object in a simulation is just an *open* surface, namely it is formed by one thin surface with zero volume. In this scenario, all the particularizations in the discussions above for a closed surface are still completely valid. However, since the magnetic field is not necessarily continuous across PEC surfaces, the unknown coefficients $J_n^{(k)}$ for $\mathbf{J}_k(\mathbf{r})$ must have, in general, independent values on the opposite sides of an open surface. There is one important exception to this rule, as proven in Subsection 3.4 of [7]. This exception appears when a scatterer body consisting of an open PEC surface is completely positioned in the interior of a homogeneous dielectric region. For example, an isolated thin PEC plate will only require to define the coefficients for $\mathbf{J}_k(\mathbf{r})$ on one side of the surface.

In our particular code implementation approach, instead of defining duplicate unknown coefficients for an open PEC surface, we simply duplicate the mesh when required, considering that the duplicate mesh comes from an independent surface. This duplication mildly increases the memory consumption as it is necessary to store two identical meshes, but, in return, is straightforward to implement in an existing MoM code. As a mesh duplication example, consider the composite geometry schematically represented in Fig. A2a where the thick black line represents an open PEC surface and the rest of the surfaces are ordinary separation surfaces between different media.

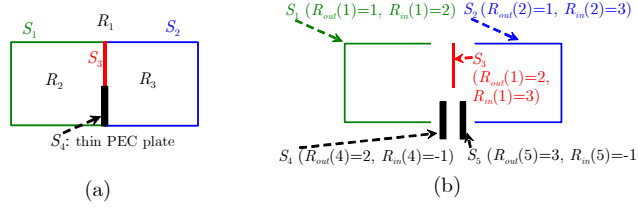

Figure A2. (a) Schematic representation of three separation surfaces and one open PEC surface. The PEC surface is represented by the thick black line. (b) Allocation of region labels for each surface in the example in (a). The duplicate meshes for the PEC plate have different labels for their outer regions: $R_{out}(4) = 2, R_{out}(5) = 3$. The gaps depicted separating the meshes in (b) are for representation clarity only.



\end{document}